\documentclass{article}
\usepackage{amsfonts}
\usepackage{amssymb}
\usepackage{amsmath}
\usepackage{amsthm}
\usepackage[english]{babel}
\begin{document}

\title{\bf On the degrees of freedom count\\ on singular phase space submanifolds}

\author{{\bf Alexey Golovnev}\\
{\small {\it Centre for Theoretical Physics, The British University in Egypt,}}\\
{\small \it El Sherouk City, Cairo 11837, Egypt}\\
{\small agolovnev@yandex.ru}}
\date{}

\maketitle

\begin{abstract}

I discuss singular loci in the phase spaces of theories which lack globally well-defined numbers of dynamical modes. This is a topic which appears quite often in the recent literature on modified gravity. In particular, there were discussions about $R^2$ gravity around Minkowski space. It is a relatively simple case, and still there were some confusions. It clearly shows that one should be very accurate when trying to understand a potentially problematic theory through perturbations around a simply looking background. At the same time, many modern teleparallel approaches are laden with even more severe issues. Therefore, it is a topic which is certainly worth carefully thinking about.

\end{abstract}

\section{Introduction}

A feasible theory of physical phenomena should better have a well-defined Cauchy problem for us to be able to make predictions which can later be compared to experiments. This is not to say that only hyperbolic equations are of any use. In the end of the day, our fixation with time development might be viewed as a mere psychological issue of trying to avoid dishonestly selling a mere description as a full explanation. At a more global level, it is of course also related with our belief in causality which might in turn be coming from our strange belief in free will. 

Nevertheless, I take the task of looking at how the models might or might not define a predictable evolution in time. To start with, ideally it means well-defined numbers of dynamical, constrained, and pure gauge modes, at least in vicinity of physically interesting backgrounds. At the same time, we often face singular loci in the phase space where such an assumption breaks down. My aim is to discuss the complications related to these situations.

Initially, this paper of mine was initiated by the fact that it has been recently noticed \cite{Lust} that a metric $f(R)$ gravity, with $f(R)=R^2$ and no linear term, has no dynamical degrees of freedom when linearised around Minkowski. It is of course immediately obvious for the usual graviton since $f^{\prime}(0)=0$, however might seem rather surprising for the scalar degree of freedom. Nonetheless, this is indeed true, in the sense of maximally using the emergent gauge freedom of that limit. I would like to show that it can easily be seen directly from the equations of motion, with no need of playing games with Lagrangians. It is very important, especially because those games are by far not always very innocent. 

Moreover, the scalar part of this "strong coupling" behaviour is actually more interpretational than dynamical. Unlike for the tensor graviton, it is not due to full disappearance of the mode from the quadratic action. It is not even a literal strong coupling of losing the kinetic energy only. It rather comes from losing a constraint. On the other hand, the same paper \cite{Lust} also reported a different number of dynamical modes, namely the scalar degree of freedom having survived the linearisation procedure, via a covariant trick \`a la St{\"u}ckelberg. This is another important topic to be discussed because the problem was due to an incorrect realisation of the trick. In the way it had been applied, it does radically change the model at hand because of fixing a gauge directly inside the action.

I will use this $R^2$ example in Section 3 and various simple toy models in Section 2 for discussion. At the same time, many modified teleparallel models exhibit more complicated bad features \cite{Tissues} which I will also briefly touch upon here and there, and more attentively in the final Section 5. Section 4 will be devoted to remarks on the Hamiltonian analysis, and in particular against the common opinion that a proper Hamiltonian for a gauge theory is the extended one.

\section{Simple examples}

I will start with the simple examples which one can construct from the usual models we have in theoretical physics. However, let me first fix the general terms.
\begin{itemize}
\item I assume that some coordinates on the spacetime manifold are chosen, and therefore all the geometric variables can be given in terms of components, i.e. spacetime functions taking values in real numbers. The total number of these functions is then the number of \underline{modes} which I now want to classify.
\item If the variables are not predictable at all, and the freedom can be parametrised by arbitrary functions of all spacetime coordinates including time, I call it gauge freedom; and I also say that the number of those independent functions is the number of \underline{\it pure gauge modes}. A non-trivial aspect here is that very often this is not a unique prescription, which particular variable to call pure gauge, and we then usually go for minimising the number of modes in the dynamical regime below.
\item In the predictable sector, I find the number of the needed Cauchy data for the prediction to be unique or almost unique, up to a possible discrete freedom due to non-linearity of equations. I say that the \underline{\it dynamical modes} are one half of this number, therefore making an analogy with the familar need of specifying the initial position and the initial velocity.
\item What is left are physical variables, for they are fixed by equations, but not dynamical, for they are uniquely fixed with no need for Cauchy data. I say that the number of \underline{\it constrained modes} is the total number of modes minus pure gauges and dynamical ones.
\end{itemize}
This is not a very precise scheme, and not always it works, but we lack a better one. Each instance of these numbers being not constant I will call \underline{strong coupling}. When something goes from being dynamical to being constrained, this is a smart term. However, I will also use it for a kind of opposite situation when something physical suddenly becomes pure gauge.

Let me also stress it once more that I take the constrained modes as physical. It goes pretty much against the commonly used jargon of calling only dynamical ones physical. One might say that it is more of a philosophical question. However, taken at the face value, the common lore leads to proclaiming that Coulomb potentials in electrodynamics or Newtonian potentials in gravity are somehow unphysical. It does not seem reasonable. From the formal perspective of an abstract dynamical model, a pure gauge quantity is something that cannot be predicted at all from the equations of motion, and therefore unphysical for the given theory, while a constrained mode is fixed by the equations, with the only difference from dynamical modes in the need of Cauchy data or lack of that, and everything which follows from the equations I call physical.

One of the brightest examples of the opposite attitude can be found in a very recent paper\footnote{Let me also use the occasion of mentioning this paper to admit that I totally do not consider the asymptotic symmetries. It is a very beautiful and interesting topic \cite{asymp} on its own, however for now I prefer to think only in local terms. Note that, for formally deriving the equations of motion, it is enough to demand stationarity of the action with respect to perturbations of strictly finite support, and then no boundary terms are ever needed.} \cite{asymp} when they (correctly) say that "it is not always straightforward to separate first-class constraints (that generate symmetries) from second-class constraints (that eliminate redundant fields)". The catch is in the parentheses. Assuming that all first-class constraints generate (gauge) symmetries is about the extended Hamiltonian framework which I will discuss later. However, claiming that the second-class ones simply eliminate redundant fields means totally neglecting part of the information which is there in the equations of motion. If we accept that all what follows from equations is physical, then these fields are not redundant, they do carry some physics. At the same time, the pure gauge ones look much more redundant, in that they do not influence anything predictable. 

\subsection{Scalar fields}

A very simple two-scalar-fields example would be
\begin{equation}
\label{1ex}
{\mathcal L}=\frac12 \left(\vphantom{\int} (\partial\phi)^2 - \phi^2 (\partial\chi)^2 \right)
\end{equation}
where $(\partial\phi)^2\equiv (\partial^{\mu}\phi)\partial_{\mu}\phi$. As long as $\phi\neq 0$, its equations
\begin{equation}
\label{1exeom}
\ddot \phi - \bigtriangleup\phi + (\partial\chi)^2 \phi=0, \qquad \phi^2 \left(\ddot\chi - \bigtriangleup\chi\right) + 2\phi \left(\dot\phi \dot\chi - (\overrightarrow{\partial\phi})\cdot(\overrightarrow{\partial\chi})\right)=0
\end{equation}
can trivially be solved for higher time derivatives (accelerations), therefore describing two dynamical degrees of freedom, with neither constraints nor gauge freedoms.

At the same time, in a singular locus of $\phi\equiv 0$, the field $\chi$ is absolutely arbitrary. Therefore, the initial data of $\phi(0,\overrightarrow{x})=\dot\phi (0,\overrightarrow{x})=0$ do not require any Cauchy data for the $\chi$ field which remains undetermined. In the intermediate case of $\phi(0,\overrightarrow{x})=0$ and $\dot\phi (0,\overrightarrow{x})\neq 0$, we would need Cauchy data for $\chi$ at some another moment of time, for at $t=0$ it would generically exhibit singular behaviour. If the field $\chi$ had a potential term, it would become infinite upon canonical normalisation in the limit of $\phi\to 0$, hence the name of strong coupling. Note that it is an infinitely strong coupling, with the initial data problem not being well-defined at all.

If we go for studying perturbation theory around $\phi=0$ and arbitrary $\chi$ background, the $\delta\chi$ perturbations do not enter the quadratic Lagrangian, while the background value of $(\partial\chi)^2$, be it zero or not, serves as a (variable) mass for $\delta\phi$. Then, in this limit, we immediately find that $\delta\phi$ is a dynamical mode, while $\delta\chi$ is, rather trivially, a pure gauge. Being totally absent from the Lagrangian, it does not "hit twice" in the Hamiltonian description, and the model has got the only (primary first-class) constraint of vanishing momentum, $\pi_{\delta\chi}=0$, which is a rare case of it indeed being a generator of the gauge transformation.

All in all, this is a very bad model (\ref{1ex}) in that its Cauchy problem and therefore the number of degrees of freedom are ill-defined. It is not only about the approximation of quadratic action (i.e. linearised theory) around the problematic background of $\phi=0$. Generically, the set of possible solutions corresponds to two dynamical degrees of freedom, however it also has the singular set which contains $\phi\equiv 0$ and absolutely arbitrary $\chi$ as solutions of the full non-linear equations (\ref{1exeom}). We might treat it as a reasonable effective field theory only as long as we somehow can always stay far away from the zero value of the field $\phi$.

Note though that sometimes these accidental constraints for momenta might be perfectly compatible with regular structure of solutions. For example, a model of
\begin{equation}
\label{2ex}
{\mathcal L}=\frac12 \phi^4 (\partial\phi)^2
\end{equation}
obviously produces a wave equation for $\phi^3$. If we attempted a Hamiltonian picture for the model (\ref{2ex}), it would get an unwanted constraint for the momentum $p$ at the locus of $\phi=0$, though with a singular term $\frac{p^2}{2\phi^4}$ in the Hamiltonian which can easily be shown to go through a finite value at a node of the wave when $\phi(x) = \sqrt[3]{\mathrm{a\ wave}}$. The invertible change of variables, from $\phi$ to $\frac13 \phi^3$, removes the singularities from the Hamiltonian analysis. Those then present themselves through the fact that the solutions in terms of initial variables $\phi(x) = \sqrt[3]{\mathrm{a\ wave}}$ do have singular derivatives at the nodes of the wave.

\subsection{Invertibility of a change of variables}

It is important to stress that the model (\ref{2ex}) above gives precisely the same results, irrespective of whether the initial variable $\phi$ or the new variable $\phi^3$ is used in the action, only due to its extreme simplicity. The equations  
$$\phi^2 \Box \phi^3 =0\quad \mathrm{and}\quad \Box \phi^3 =0$$
do practically have the same set of solutions because any derivative of an identical zero is also zero, so that an option of $\phi^2=0$ does not add anything new to the case of $\Box \phi^3 =0$. At the same time, this change of variables $\phi\leftrightarrow\phi^3$, despite being invertible (in the realm of continuous functions), modifies the variational procedure at the locus of $\phi=0$ which comes in terms of an extra factor of $\phi^2$ by which the two equations differ from each other. 

In more complicated models it might change the physics. For example, such changes of variables have been used for producing new mimetic-like solutions \cite{Vik1, Vik2}. Namely, they represent the physical metric as a disformal transformation of another one
$$g_{\mu\nu}=C(\phi, {\tilde X})\cdot {\tilde g}_{\mu\nu} + D(\phi, {\tilde X})\cdot  (\partial_{\mu}\phi)\partial_{\nu}\phi, \qquad {\tilde X}\equiv {\tilde g}^{\mu\nu} (\partial_{\mu}\phi)\partial_{\nu}\phi$$
which is then treated as a new fundamental variable together with the auxiliary scalar field. One can easily see \cite{antiV} that the metric transformation is invertible if and only if a simple scalar relation $X=X({\tilde X})$ is. Then it is not hard to show \cite{DeR} that invertible cases do not produce anything new and can be called "veiled general relativity (GR)", while otherwise a particular mode of the physical metric gets represented by the scalar field only, and due to the extra derivatives in the action functional, it yields more general equations of mimetic gravity \cite{mimet,memim}.

The classical mimetic gravity \cite{mimet} is about $X\equiv 1$, and therefore the transformation is never invertible. The new idea of the papers \cite{Vik1, Vik2} was in using a transformation with $\frac{dX}{d{\tilde X}}$ vanishing only for some particular value of $\tilde X$. At the price of an ill-defined number of degrees of freedom, one can get \cite{Vik1, Vik2} the mimetic type solutions on top of the pure GR for these special configurations of the scalar field. This is an example of one-to-one changes of variables being able to modify the model. With the derivative vanishing at an isolated point, the function $X(\tilde X)$ can still be bijective and invertible in the class of continuous functions, but not continuously differentiable ones. 

In my opinion \cite{antiV}, it is a very problematic procedure, in the sense of producing an ill-posed Cauchy problem. However, it is definitely an interesting and non-trivial aspect of singular behaviours. Note also that effects of various higher-derivative generalisations of such changes of variables inside the action are being actively studied nowadays \cite{Babich}.

\subsection{An interlude: the Proca field}

Now we switch to vector fields which are by far more interesting. First, let us recall the standard Proca, i.e. massive vector field,
\begin{equation}
\label{stvec}
{\mathcal L}=-\frac14 F_{\mu\nu} F^{\mu\nu} + \frac{m^2}{2} A_{\mu}A^{\mu}.
\end{equation}
Its equation of motion, 
\begin{equation}
\label{standeq}
\partial_{\nu}F^{\nu\mu} +m^2 A^{\mu}=0,
\end{equation}
has a costraint in the temporal component and also an obvious consequence (if $m\neq 0$) of $\partial_{\mu} A^{\mu}=0$ which is very convenient for using in the spatial part of equation in order to exclude $\dot A_0$. All in all, the equations of motion (\ref{standeq}) can be written as
\begin{equation}
\label{stveceq}
\left(\Box +m^2\right) A_i = 0, \qquad \left(\bigtriangleup - m^2\right) A_0 = \partial_i \dot A_i, \qquad \dot A_0 = \partial_i A_i.
\end{equation}
The system is redundant, for as long as we are in a domain of invertibility of the $\bigtriangleup - m^2$ operator, the third equation (divergencelessness) can be derived by time-differentiating the second one (the temporal component) and using the first. With the same accuracy, one can also rewrite the first equation for all vector field components as 
$$\left(\Box +m^2\right) A_{\mu} = 0.$$

Assuming that the field is everywhere smooth and decays at spatial infinity, and the mass is normal, i.e. $m^2>0$, the operator $\bigtriangleup - m^2$ is indeed invertible (and negative definite). It means that we have got strictly three dynamical degrees of freedom (three pairs of Cauchy data for $A_i$), and then $A_0$ is a constrained variable uniquely determined for any solution of $A_i$. In the case of a tachyonic mass,  $m^2<0$, according to the equations (\ref{stveceq}), the modes with $k^2=-m^2$ do have the longitudinal part $\partial_i A_i$ constant in time and uniquely determining the velocity $\dot A_0$. Therefore, it is still a pair of Cauchy data for the longitudinal and temporal components, the initial values of $\partial_i A_i$ and $A_0$, which are free to be chosen. All in all, it can still be counted as three dynamical degrees of freedom and one constrained mode.

In the same assumption of the field tending to zero at spatial infinity, one can uniquely decompose its spatial components into transversal and longitudinal modes,
$$A_i=A^{\mathcal T}_i+ \partial_i\varphi, \qquad \partial_i A^{\mathcal T}_i \equiv 0,$$
with the transverse part obviously having two independent components. In this case, the equations (\ref{stveceq}) can be rewritten as
\begin{equation}
\label{stvhaml}
\left(\Box +m^2\right) A^{\mathcal T}_i = 0, \qquad \dot A_0 = \bigtriangleup \varphi, \qquad  \bigtriangleup \dot \varphi = \left(\bigtriangleup - m^2\right) A_0.
\end{equation}
Therefore, we have the two familiar degrees of freedom in the transversal polarisations, with the third dynamical mode presented as a first-order system for $A_0$ and $\bigtriangleup\varphi$. This is true of either sign of $m^2$, as well as for the Lorenz gauge of the gauge theory at $m=0$.

Note that if we restrict our consideration to bounded fields vanishing at spatial infinity, it's possible to use $\bigtriangleup\left(\partial_{\mu}F^{\mu i} +m^2 A^i\right)=0$ instead of the spatial equation. Then substituting $\bigtriangleup A_0  = \partial_i \dot A_i + m^2A_0$ from the temporal equation, we get
\begin{equation}
\label{funveceq}
\Box (\bigtriangleup\delta_{ij}-\partial_i \partial_j)A_j + m^2 (\bigtriangleup A_i - \partial_i \dot A_0) =0.
\end{equation}
In any case, the transverse components must satisfy a simple wave equation. If $m=0$, this is the only dynamical information, while the temporal and longitudinal sector has one constraint of $\bigtriangleup A_0  = \partial_i \dot A_i$ and one gauge freedom. If $m\neq 0$, divergence of the equation (\ref{funveceq}) imposes vanishing of the 4-vector divergence, and we easily come to the three dynamical degrees of freedom of the equations (\ref{stvhaml}).

As a final remark, one can immediately employ the condition of divergencelessness of the vector $A_{\mu}$, i.e. $\partial_{\mu} A^{\mu}=0$, by reparametrising it via
\begin{equation}
\label{1-3forms}
A^{\mu}=\epsilon^{\mu\nu\alpha\beta}\partial_{\nu}s_{\alpha\beta}, \qquad s_{\alpha\beta}=-s_{\beta\alpha}.
\end{equation}
Such a vector (\ref{1-3forms}) obviously has zero divergence. At the same time, the antisymmetric matrix $s$ of a 2-form possesses six algebraically independent components, while a new gauge freedom of shifting it by differentials of non-exact 1-forms subtracts three modes and leaves us with the correct number of three modes in $A_{\mu}$. It can also be understood as a statement that, in case of a trivial spacetime topology, a 1-form which corresponds to a divergenceless vector has got an exact 3-form as its Hodge dual.

Substituting it (\ref{1-3forms}) into the equation (\ref{standeq}), we get a 3rd order equation
$$\epsilon^{\mu\nu\alpha\beta} \partial_{\nu} (\Box + m^2) s_{\alpha\beta}=0$$
which, modulo the new gauge freedom, shows the same wave equation for all three dynamical modes in terms of $A_{\mu}$. Additional exact contributions to the 2-form $s$ are pure gauge of this representation and are not fixed anyhow. However, putting this reparametrisation (\ref{1-3forms}) directly into the action (\ref{stvec}) would lead us to 4th order equations of motion thus changing the model at hand.

\subsection{The St{\"u}ckelberg trick}

Once we have introduced a mass term, the vector is no longer $U(1)$ gauge invariant. The idea of the St{\"u}ckelberg trick is that instead of one vector field we may work with one vector and one scalar,
\begin{equation}
\label{stueck}
A_{\mu} \longrightarrow A_{\mu} + \partial_{\mu} \phi,
\end{equation}
such that the new system enjoys a gauge symmetry of
$$\phi \longrightarrow \phi+\chi, \qquad A_{\mu} \longrightarrow A_{\mu} - \partial_{\mu} \chi.$$
This is, of course, an artificially introduced $U(1)$ symmetry, while the gauge invariant variables correspond to the initial model we've started with. Despite the derivative in the definition (\ref{stueck}), all gauge invariant variations for the initial action (for the old $A_{\mu}$) are now given by variations of the new $A_{\mu}$, with no influence on boundary conditions.

The action (\ref{stvec}) then takes the form of
$${\mathcal L}=-\frac14 F_{\mu\nu} F^{\mu\nu} + \frac{m^2}{2} \left(\vphantom{\frac12} A_{\mu}A^{\mu}+ 2A^{\mu}\partial_{\mu}\phi + (\partial^{\mu}\phi)\partial_{\mu}\phi\right)$$
with the equations of motion
$$\partial_{\nu}F^{\nu\mu} +m^2\left( A^{\mu} +\partial^{\mu}\phi\right) =0, \qquad \Box\phi + \partial_{\mu} A^{\mu}=0.$$
Initially, a vector of the gradient shape did not allow for any non-trivial solution. Now the scalar equation looks like $\phi$ being dynamical. This is because of the higher derivative with which it comes into the action. However, if $A_{\mu}=0$, then the vector equation demands that $\phi$ is constant, i.e. pure gauge not influencing the gauge invariant combination. 

All in all, the physical vector $A_{\mu} + \partial_{\mu} \phi$ has got absolutely the same dynamics as the vector of the initial non-covariant model. The new scalar equation is simply about its divergencelessness. As the paper \cite{Lust} correctly says referring to the classical review on massive gravity \cite{Hinter}, the St{\"u}ckelberg trick is not a reparametrisation but rather an introduction of a new variable. This is true. The key is in the construction of a gauge freedom which leaves the invariant quantities the same as before the trick.

However, using this trick, they had got incorrect results \cite{Lust} on the number of degrees of freedom, i.e. more modes than in reality, for example the scalar dynamical mode in the purely $R^2$ gravity linearised around Minkowski.  The reason is that it had been used in an incorrect way. Namely, instead of introducing a new variable, they \underline{did a reparametrisation} of
\begin{equation}
\label{unstueck}
A_{\mu}=A^T_{\mu} + \partial_{\mu} \phi, \qquad \partial^{\mu} A^T_{\mu}\equiv 0.
\end{equation}
This reparametrisation is a bit redundant. One can shift the variables by $\partial_{\mu}\chi$ and $-\chi$ as long as the change satisfies an equation of $\Box\chi=0$. In case of separating a 3D-longitudinal mode, we may fix such redundancy by spatial boundary conditions, but we always treat the time differently. It can also be thought of as an (incomplete) gauge fixing \cite{Lust} of the model defined by the trick (\ref{stueck}). This is all right at the level of equations of motion. However, this gauge fixing right inside the action drastically changes the model.

To be precise, the parametrisation (\ref{unstueck}) does not change anything in the variation of 4D-transverse modes. However, in the 4D-longitudinal part we have got $\partial_{\mu}\delta\phi$ instead of $\delta A_{\mu}$. This extra derivative in the action produces a more general equation than it was before "fixing the gauge", due to changes in how the boundary conditions work \cite{antiV,memim}. The variation of the action (\ref{stvec}) becomes
$$\delta S = \int d^4 x \left(\left(\partial_{\nu}F^{\nu\mu} +m^2 {A^T}^{\mu}\right)\delta A^T_{\mu} - m^2 (\Box \phi)\delta\phi \right).$$
Only the divergenceless part of the coefficient in front of $\delta A^T_{\mu} $ must vanish, but it automatically has this property, and this already corresponds to the three degrees of freedom of a massive vector field. The second term in the variation is one more, scalar degree of freedom. That's how a gauge fixing inside the action, or a reparametrisation with derivatives, can drastically change a dynamical model.

One can produce dynamics out of nothing. Consider a trivial action functional of the Lagrangian density ${\mathcal L} = A_{\mu}A^{\mu}$ with neither dynamical degrees of freedom nor gauge freedom. All four modes are constrained, $A_{\mu}=0$. If we do a St{\"u}ckelberg trick $A_{\mu}\longrightarrow A_{\mu}+\partial_{\mu}\phi$, an obvious gauge freedom appears. However, either in the equations of motion or in the action, it leads to the very same equation for the physical variable $A_{\mu}+\partial_{\mu}\phi=0$ as before. At the same time, if one goes for a reparametrisation, or a gauge-fixed St{\"u}ckelberg trick, $A_{\mu}=A^T_{\mu}+\partial_{\mu}\phi$ with $\partial_{\mu}{A^T}^{\mu}\equiv 0$, then it is all right at the level of equations, but inside the action it leads to more general equations, $A^T_{\mu}=0$ and $\square\phi=0$, which describe more freedom than before, even for the physical variable $A^T_{\mu}+\partial_{\mu}\phi$.

One can go this way as far as (s)he wants. Even with a scalar theory, we can do a trick of $\phi \longrightarrow \phi + \Box\psi$. The new variables enjoy the gauge symmetry of $\phi\longrightarrow\phi+\Box\chi$ and $\psi\longrightarrow\psi-\chi$. If the initial model had an equation ${\mathcal F}[\phi]=0$, then the new one gets ${\mathcal F}[\phi+\Box\psi]=0$ and $\Box{\mathcal F}[\phi+\Box\psi]=0$. We've got a higher derivative model which is still equivalent to the initial one. Indeed, the first equation is precisely the same equation as it used to be, just acting on the gauge invariant variable, while the second equation follows from the first one, though in itself it enjoys many more solutions. Had I taken a crazy idea of fixing the $\phi=0$ gauge right inside the action, it would have given me only the second equation, $\Box{\mathcal F}[\Box\psi]=0$, which is higher order even for the gauge invariant information, due to the variational prescription having been modified. Basically, the Lorenz-type gauges inside an action \cite{Lust} commit the same crime. More elaborate examples of this very strange procedure can, for example, be found\footnote{Just to be clear, I do not recommend taking it seriously.} in Ref. \cite{recent}.

\subsection{Proca field with a variable mass}

Since massive and massless vector fields are different in their numbers of dynamical modes, a model of variable mass,
\begin{equation}
\label{3ex}
{\mathcal L}=-\frac14 F_{\mu\nu} F^{\mu\nu} + \frac12 (\partial\psi)^2 + \psi A_{\mu}A^{\mu} + c \psi,
\end{equation}
is then obviously singular at the locus of $\psi=0$. Away from it, there is no gauge freedom, one constrained variable ($A_0$) and four dynamical modes (the spatial components $A_i$ of the vector and the scalar $\psi$). The field $\psi$ gives an effective mass-squared $2\psi$ to the vector field $A_{\mu}$, and also has a linear potential $c\psi$ where $c$ is a constant number. Basically, it is a canonical free field with a source term.

The equations of motion are
\begin{equation}
\label{3exeq}
\partial_{\nu} F^{\nu\mu} + 2 \psi A^{\mu} =0, \qquad \Box\psi = A_{\mu}A^{\mu} + c.
\end{equation}
Suppose that $c>0$ and take the coordinates $(t,x,y,z)$, then there exists a solution
$$A_{\mu}=(0,\ \sqrt{c},\ 0,\ 0), \qquad \psi=0$$
to the equations (\ref{3exeq}). Either by considering the quadratic action or directly perturbing the equations, one can find
$$\partial_{\mu}\delta F^{\mu\nu} = 2\sqrt{c} \delta^{\nu}_1 \delta\psi , \qquad \Box\delta\psi =-2\sqrt{c} \delta A_1$$
for linearised perturbations around. We immediately see, by taking divergence of the vector equation, that $\delta\psi$ cannot depend on the $x$ coordinate. Therefore, the scalar field enjoys a reduced freedom of Cauchy data, that of $\delta\psi(0,x,y,z)=f_1(y,z)$ and $\dot\delta\psi(0,x,y,z)=f_2(y,z)$ only, while the vector field gets the usual gauge freedom in a reduced form of $\delta A_{\mu}\longrightarrow \delta A_{\mu}+ \partial_{\mu}\chi(t,y,z)$. This is somewhat reminiscent of an extra mode with an incomplete Cauchy freedom around non-trivial Minkowski background of $f(\mathbb T)$ gravity \cite{meM}.

To a certain extent, the ugly amount of gauge freedom is an artifact of the linear approximation. Indeed, if we restrict ourselves to solutions with $\psi=0$ everywhere, then formally we simply have a massless vector field $\delta A_{\mu}$ with a gauge condition of $\delta A_1=0$ on top. This gauge fixing being incomplete leaves the remnant symmetry parametrised by $\chi(t,y,z)$. The problem is that it is not just a gauge choice but a requirement imposed by equations. Going to higher orders, we easily see that
$$A_{\mu}=(0,\ \sqrt{c},\ 0,\ 0) + \partial_{\mu}\chi \qquad \mathrm{with}\qquad (\partial_{\mu} \chi)^2 = 2\sqrt{c} \partial_x \chi$$
solves the full equations with $\psi=0$. Therefore, at higher orders, the longitudinal mode becomes dynamical.

More precisely, a combination with $\psi\equiv 0$ demands that the vector field $\delta A_{\mu}$ satisfies electrodynamic equations in vacuum $\partial_{\mu}\delta F^{\mu\nu}=0$ with the gauge condition of $\delta A_0 =\pm\sqrt{ 2\sqrt{c} \delta A_1 + \delta A^2_i}$. Analytically, it works like three degrees of freedom, though with a singular behaviour at the limit of weak perturbations: for the spatial components of order small $\epsilon$, the temporal one goes like $\sqrt{\epsilon}$. In this sector, it is somewhat similar to the situation with the scalar field (\ref{2ex}) where the analytical dynamics is under control. With a big portion of wishful thinking, people sometimes hope that the strong coupling issues of $f(\mathbb T)$ gravity might also get a resolution in going beyond the linear perturbation theory.

However, the dynamical problem (\ref{3exeq}) is ill-posed, and this is not only an issue of unreliable approximations. Generically, away from zero value of the field $\psi$, there is no gauge freedom while the initial data have two restrictions: an analogue of the Gau\ss{}  law $(\bigtriangleup- 2\psi) A_0 = \partial_i \dot A_i$ and the divergence of the vector equation $\partial_0 (\psi A_0)=\partial_i (\psi A_i)$ which altogether make one degree of freedom non-dynamical. The former condition is unavoidable, while the latter relation does not restrict the initial data for the vector field anyhow if the initial $\psi(0,\overrightarrow x)$ and $\dot\psi(0,\overrightarrow x)$ are taken to vanish. And indeed, there is nothing in the model to stop the vector field from sourcing appearance of non-zero $\psi$ at $t>0$. Above we simply adopted yet another constraint for making $\Box\psi$ vanish, too.

On the other hand, since the $A_{\mu}$ field is assumed to be physical in itself with no gauge freedom on top, one can look at spatially homogeneous configurations. If $\partial_i A_{\mu}=0$, we see (\ref{3exeq}) that either $\psi=0$ or $A_0=0$. In the former case, one gets three dynamical modes in $A_i$ and a constraint for $A_0$, while in the latter case four dynamical variables, $A_i$ and $\psi$, come about.

Any hope for improving the dynamical properties in such situations cannot then lie in taking higher orders of perturbation theory. The very model must then be changed. It's a possible option in the spirit of effective field theory \cite{strongeft}. The model at hand might simply be some unsuccessful approximation to a more fundamental action. However, then the first question is whether a healthy and viable model exists at all, in a given class of theories. And even if the answer is affirmative, if the singular loci of the initial approximation were found in the physically interesting regions, then how can we ever trust anything we have seen there. Naively estimating the strong coupling scales \cite{1strong} is just a possible way to close our eyes at the mathematical problem, for some time at least. Moreover, philosophically, reliance on adding new interaction terms when the initial coupling was strong is based on an idea that all the relevant variables stayed physical. It can be applied to cases of a kinetic term getting a vanishing coefficient, therefore producing an infinite factor in an interaction term upon canonical normalisation, but it cannot work like that for an accidental gauge symmetry though.

Finally, a very simple example of a strong coupling, in terms of accidental gauge symmetry, can be seen in the action of
\begin{equation}
\label{4ex}
{\mathcal L}=-\frac14 F_{\mu\nu} F^{\mu\nu} + \frac12 (\partial\psi)^2 + \psi^2 A_{\mu}A^{\mu}.
\end{equation}
The locus of $\psi=0$ corresponds to a massless vector field. Initial data of $\psi(0,\overrightarrow x)=\dot\psi(0,\overrightarrow x)=0$ are a separate world with gauge freedom in the full equations. It would be so even if we took a non-Abelian field which has its own equations non-linear. Going across this world is generically not smooth at all. And precisely like it was with the scalar field model (\ref{1ex}), the model (\ref{4ex}) is sick at the full non-perturbative level, not only in linear approximations. Generically, all the variables are physical, with 4 dynamical modes and one constrained quantity ($A_0$), however one can also have solutions with $\psi\equiv 0$ and one totally unpredictable mode in the vector field.

\subsection{Non-linear self-interacting Proca}

For a more intricate example of a potentially unstable nature of a dynamical model, one can take a generalised Proca field,
\begin{equation}
\label{Proca}
{\mathcal L}=-\frac14 F_{\mu\nu} F^{\mu\nu} + V(A_{\mu}A^{\mu})
\end{equation}
with the equation of motion
\begin{equation}
\label{eomProca}
\partial_{\mu} F^{\mu\nu} + 2 V^{\prime} A^{\nu}=0
\end{equation}
and $V$ being minus the potential energy. The generalisation (\ref{Proca}, \ref{eomProca}) compared to the classical Proca action (\ref{stvec}) is in a more general potential than the simple mass term. I don't use the "generalised Proca" collocation in the title of the subsection because nowadays it is commonly used in a different, much more elaborate meaning \cite{Lavvec1, Lavvec2}.

The main point for us now is that generically, away from $V^{\prime}=0$, we have no gauge freedom, one constrained mode $A_0$ and three dynamical modes $A_i$. Possible $V^{\prime}=0$ loci of non-vanishing $A_{\mu}$ do have a slightly more complicated structure of symmetry breaking, in the perturbations along the background field. Indeed, the quadratic Lagrangian around it gets a term of $2V^{\prime\prime}(A_{\mu} \delta A^{\mu})^2$, with the $4V^{\prime\prime}A^{\nu}A_{\mu}\cdot \delta A^{\mu}$ term in the equation of motion. For a purely spatial background field, we get a story of incomplete gauge freedom similar to what we have seen in the previous example (\ref{3ex}).

An accidental gauge symmetry fully appears for $V^{\prime}=0$  when either  also $V^{\prime\prime}=0$ or the background vector vanishes, $A_{\mu}=0$. As for the simplest case, I take a model of $V(A_{\mu}A^{\mu})=-\frac14(A_{\mu}A^{\mu})^2  $
\begin{equation}
\label{exProca}
{\mathcal L}=-\frac14 \left(\vphantom{\int} F_{\mu\nu} F^{\mu\nu}  + (A_{\mu}A^{\mu})^2 \right)
\end{equation}
around the $A_{\mu}=0$ vacuum. The equation of motion,
\begin{equation}
\label{eomex}
\partial_{\mu} F^{\mu\nu} -A_{\mu} A^{\mu} A^{\nu}=0,
\end{equation}
generically gives a well-defined evolution of the vector field, with the needed Cauchy data for three dynamical degrees of freedom. Due to non-linearity, there can be different regimes. For example, a spatially homogeneous field can either behave as a non-linear purely spatial field, or as a massless null vector. Generically, both regimes have three degrees of freedom, also for linear perturbations around these backgrounds except the vacuum case of $A_{\mu}=0$. However, around the vacuum, the linearised theory acquires the usual $U(1)$ gauge freedom.

At the same time, this model (\ref{exProca}) is more intricate in the types of its bad behaviour. One can look at the common sources of restrictions on Cauchy data for the equation of motion (\ref{eomex}): its temporal component
\begin{equation}
\label{fun1}
\bigtriangleup A_0 - \partial_i \dot A_i + (A_0^2 - A^2_j) A_0 =0
\end{equation}
and its divergence
\begin{equation}
\label{fun2}
(3A^2_0 - A^2_j)\dot A_0 - 2A_0 A_j \dot A_j -\partial_i \left((A^2_0 - A^2_j)A_i\right)=0.
\end{equation}
Generically, the initial velocities of the temporal and longitudinal modes get constrained thus removing one dynamical degree of freedom. However, if the initial values are chosen such that $3A^2_0 = A^2_j$, the velocity $\dot A_0$ drops off. Then, despite the fact that the variable $A_0$ is fully detemined by the spatial components, those still have yet another constraint among themselves. The first equation (\ref{fun1}) can be solved for $A_0$ as a function of $A_i$ and $\dot A_i$, then the second equation (\ref{fun2}) gives a new constraint on the purely spatial Cauchy data, as long as $3A^2_0 - A^2_j =0$ at the initial time. Again, it is a locus in the configuration space at which the Cauchy problem gets ill-defined. This can be easier seen in the Hamiltonian language to which I will come later.

\section{$R^2$ gravity around Minkowski}

Among the modified gravity models, those of $f(R)$ type are very well studied and have got rather good perspectives, both academically and phenomenologically. Whenever $f^{\prime}\neq 0$ and $f^{\prime\prime}\neq 0$, they can be translated into a model of GR with a canonical scalar field, by a conformal transformation of the metric. Normal or ghost nature of both types of dynamical modes depends on the signs of the function $f$ derivatives. At the same time, at the locus of $f^{\prime}=0$, the standard general relativistic contribution obviously disappears from the linearised theory, with the only quantities which remain coming from the total derivative part of the scalar curvature responsible for the new dynamical scalar.

Recently a model of "pure quadratic" gravity
\begin{equation}
\label{R-squared}
S= \int d^4 x \sqrt{-g}\cdot R^2
\end{equation}
was discussed concerning its number of degrees of freedom \cite{Lust}. Obviously, it always enjoys $f^{\prime\prime}\neq 0$. However, precisely the Minkowski space, and every other vacuum GR solution, or even an arbitrary space of $R=0$ are on its problematic phase space surface, for $f^{\prime}(0)=0$. The usual gravitons immediately disappear from the linearised theory. In the present paper I do call it strong coupling, even though it is not the usual strong coupling in the literal meaning of the word which would make them constrained; it is rather an accidental gauge symmetry totally removing the vector and tensor modes from the physical sector. At the same time, a probably less expected claim \cite{Lust} is that the scalar dynamical mode is no longer alive either.

At first glance, it might sound surprising. Indeed, having easily found that
$$g^{\mu\nu}\delta R_{\mu\nu} = \left(\bigtriangledown^{\mu}\bigtriangledown^{\nu}-g^{\mu\nu}\square\right) \delta g_{\mu\nu} = \left(g_{\mu\nu}\square - \bigtriangledown_{\mu}\bigtriangledown_{\nu} \right) \delta g^{\mu\nu},$$
one immediately arrives at the equation of motion
\begin{equation}
\label{RsqEq}
2R\cdot R_{\mu\nu} - \frac12 R^2\cdot g_{\mu\nu} +2 \left(g_{\mu\nu}\square - \bigtriangledown_{\mu}\bigtriangledown_{\nu}\right)R =0
\end{equation}
for the action functional (\ref{R-squared}). Its trace implies
\begin{equation}
\label{trtr}
\square R =0
\end{equation}
which is basically the new scalar equation of motion, instead of the usual $R=0$ vacuum result, and then the equation (\ref{RsqEq}) describes the two remaining polarisations of the graviton. Obviously, the trace part of equations (\ref{trtr})  survives taking the linear limit around the Minkowski metric. However, even a stronger requirement of $\partial_{\mu}\partial_{\nu}R =0$ does so (\ref{RsqEq}). 

Intuitively, the last remark already shows that the scalar mode must also go away because local perturbations of the scalar curvature have been killed. In the paper \cite{Lust}, more technical explanations for this fact are given. However, they present two methods of analysis, one of which indeed gives this answer while another one still shows a scalar mode in the spectrum. The reason is that they intensively use derivative substitutions into the action which generically change the theory at hand. Therefore, I describe everything what happens directly at the level of equations of motion, and also show how the derivative changes of variables modify the theory leading the Authors of Ref. \cite{Lust} to the contradictory results. Basically, it all has already been illustrated on simple examples in the previous section.

\subsection{The weak field limit of pure $R^2$ gravity}

To analyse the weak gravity limit, it is enough to simply take
$$g_{\mu\nu}=\eta_{\mu\nu}+h_{\mu\nu}$$
with $h_{\mu\nu}$ being an infinitesimally  small quantity. It implies
\begin{equation}
\label{R}
R=\left(\partial_{\mu}\partial_{\nu}-\eta_{\mu\nu}\square\right) h^{\mu\nu} + {\mathcal O}(h^2)
\end{equation}
and, with the same ${\mathcal O}(h^2)$ accuracy, it also reduces the equation (\ref{RsqEq}) to
\begin{equation}
\label{weakeq}
\left(g_{\mu\nu}\square - \partial_{\mu}\partial_{\nu}\right) R=0 \qquad \mathrm{or\ simply} \qquad  \partial_{\mu}\partial_{\nu} R=0.
\end{equation}

Looking at the expression (\ref{R}) for the scalar curvature, we see that all the vector and tensor (divergenceless and traceless) perturbations are pure gauge in this limit. At the same time, the usual diffeomorphism symmetry allows us to remove one half of the scalar quantities and take the Newtonian gauge of
\begin{equation}
\label{Nmetr}
g_{\mu\nu}dx^{\mu} dx^{\nu} = (1+2\phi) dt^2 - (1-2\psi)\delta_{ij} dx^i dx^j
\end{equation}
which means that the only non-zero perturbation quantities are $h_{00}=2\phi$ and $h_{ij}=2\psi\delta_{ij}$. We then get
\begin{equation}
\label{Rpot}
R=6\square\psi + 2\bigtriangleup \left(\phi + \psi\right). 
\end{equation}
The nice property of this gauge is that those potentials numerically coincide with the gauge-invariant Bardeen ones.

Taking the perturbation theory approach, as well as following the Ref. \cite{Lust}, for the time being let's take the equation (\ref{weakeq}) simply as
$$R=0.$$
In this case, the equation of motion (\ref{weakeq},\ref{Rpot}) gets reduced to\footnote{Note that the corresponding result of Ref. \cite{Lust} is different by the relative sign of the two terms in the formula (\ref{redeq}). This is due to the opposite signature convention which reverts the sign in the definition of the $\square$ operator. I am grateful to the Authors of Ref. \cite{Lust} for pointing this out to me, for initially I wrongly thought that there was some mistake of theirs.}
\begin{equation}
\label{redeq}
3\square\psi + \bigtriangleup \left(\phi + \psi\right) =0.
\end{equation}
Since there is only one scalar equation left, there cannot be two different physical variables. One possible interpretation is that $\psi$ in itself is a pure gauge mode, while the only physical variable, $\phi$ or $\psi+\phi$, is constrained.  Doing it another way around could look like having a dynamical mode in $\psi$. However, it can also then be treated as nothing but remaining gauge freedom.

The paper \cite{Lust} observes that, upon substitution of the solution for $\psi+\phi$ into the action, the latter turns out to be identically zero, in order to claim that there is nothing more in the physical spectrum. This fact is trivially true because the equation is $R=0$ while the Lagrangian is $R^2$.  However, I would not use it as a proof that there is nothing else in the model. Substitution of a partial solution directly into the action might radically change its properties. The real proof is that, in the perturbative treatment, all the equations got reduced to only this one (\ref{redeq}).

Perturbatively, one can uniquely find the field $\phi$ in terms of $\psi$. If to take the model of 
\begin{equation}
\label{R2linact}
S=\int d^4x \cdot \left(\vphantom{\int}\left(\partial_{\mu}\partial_{\nu}-\eta_{\mu\nu}\square\right) h^{\mu\nu} \right)^2
\end{equation}
as a fundamental one, we have to solve the equation (\ref{weakeq}) as
\begin{equation*}
 \bigtriangleup \left(\phi + \psi\right) =  - 3\square\psi + b + c_0 t + c_i x^i
\end{equation*}
with constant numbers $b$ and $c_{\mu}$, which means a remaining full freedom of a spatially harmonic function with arbitrary dependence on time, and also one and three halves global degrees of freedom. 

More precisely, without relying on cosmological perturbations' parametrisation, the theory (\ref{R2linact}) has got an equation of motion
$$\partial_{\alpha}\partial_{\beta}\left(\partial_{\mu}\partial_{\nu}-\eta_{\mu\nu}\square\right) h^{\mu\nu} =0$$
which can be solved as $\left(\partial_{\mu}\partial_{\nu}-\eta_{\mu\nu}\square\right) h^{\mu\nu} =b + c_0 t + c_i x^i$ and therefore implies that the $\bigtriangleup h_{00}$ quantity can be given in terms of other metric components, with an addition of $ b + c_0 t + c_i x^i$. Modulo the global addition and the freedom of a harmonic function, it can be taken as almost everything pure gauge and one constrained physical variable.

On the other hand, if we remember that the linearised model is just an approximation, the unpredictability of a variable is not necessarily a true gauge freedom. This is what happens here. However, while the tensor and vector perturbations have totally disappeared from the quadratic action, the story of the scalar mode is a more intricate one. If we take a non-trivial spatially flat vacuum cosmology, then $R\neq 0$ and the extra scalar field takes the form of an ideal fluid in GR, therefore requiring $\phi=\psi$. If the limit of Minkowski space is taken while keeping this condition, then we have got a scalar degree of freedom. In other words, the count of degrees of freedom has a serious interpretational aspect here. There is no vanishing of the corresponding eigenvalue of the kinetic matrix. We simply have lost a constraint which used to make the initial data physical, in the meaning of influencing  predictable outcomes.

Note though that the locus of $R=0$ is non-perturbatively problematic. Indeed, any metric with $R\equiv 0$, i.e. subject to just one scalar equation, solves the full non-linear system of equations (\ref{RsqEq}).

\subsection{A parody of the St{\"u}ckelberg trick}

Recall also that the paper \cite{Lust} reported a contradictory result, namely that they still saw a scalar mode in the pure $R^2$ theory linearised around Minkowski in what was called a manifestly covariant or \`a la St{\"u}ckelberg approach. In all other cases of using this approach, the problem was simply in fixing a gauge right inside the action. Now it is slightly more problematic. They represented the metric perturbation in the form which can be taken roughly as (more precisely, see below)
\begin{equation}
\label{strpar}
h_{\mu\nu} =\quad \mathrm{vector\ and \ tensor\ contributions } \quad + \partial_{\mu}\partial_{\nu} \Sigma + \eta_{\mu\nu} \Psi.
\end{equation}
As has been mentioned before, even if there were some gauge symmetry considerations behind, it would definitely not be the standard St{\"u}ckelberg procedure which introduces new variables without restricting the old ones (by zero trace or by zero divergence). However, in this case, it is even worse. With the classical St{\"u}ckelberg trick, the gauge-invariant combination is immediately obvious, while ambiguity of the spin formalism is much less transparent.

We can immediately calculate the (linearised) Ricci scalar (\ref{R}) to be $R=-3\square\Psi$. If, in perturbation theory again, we treat the equation (\ref{weakeq}) as $R=0$, we get a dynamical mode of $\square\Psi=0$. The reason is that this representation (\ref{strpar}), taken \cite{Lust} from a previous paper \cite{inSt}, is very ambiguous\footnote{One might worry that this parametrisation looks very similar to cosmological perturbation theory. However, the latter works in terms of spatial derivatives only, and ambiguities of solving equations for Laplacians are taken care of by boundary conditions at spatial infinity which is absolutely reasonable when in perturbation theory.}. Let's, for example, take the metric (\ref{Nmetr}) which did not lead to any dynamical mode and find the trace and the double divergence of its perturbation for presenting it in the shape of (\ref{strpar}). We get 
$$\square\Sigma + 4\Psi=2\phi - 6\psi\qquad \mathrm{and} \qquad \square^2\Sigma +\square\Psi=2\ddot\phi +2\bigtriangleup\psi,$$
and can express $-3\square\Psi$ in terms of $\phi$ and $\psi$. In other words, the two Cauchy data for $\Psi$ simply correspond to a freedom of choosing a presentation of the metric in the shape of the formula (\ref{strpar}).

At the same time, the action (\ref{R2linact}) turns into $9\int d^4 x \cdot \left(\square\Psi\right)^2$ and produces even a 4-th order equation of motion $\square^2 \Psi=0$. In this case, the field $\Psi$ was not given any extra derivatives in the action, however the operator of $\partial_{\mu}\partial_{\nu}$ had effectively been removed from it producing $\square R=0$ instead of $ \partial_{\mu}\partial_{\nu} R=0$. Therefore, we get more solutions than the initial model used to have. It is yet another example of problems with substitutions right into the action.

Finally, to be precise, the scalar part of the metric perturbation (\ref{strpar}) was taken \cite{Lust, inSt} in a bit different form of
$$ \left(\partial_{\mu}\partial_{\nu} -  \frac14 \eta_{\mu\nu}\square\right) \mu + \frac14 \eta_{\mu\nu} \lambda = \partial_{\mu}\partial_{\nu} \mu + \frac14 \eta_{\mu\nu} \left( \lambda  -   \square\mu\right),$$
I guess, with motivations from spin projector formalism so that the derivative part be traceless, which has of course given them a Lagrangian density proportional to $\left( \lambda  -   \square\mu\right)^2$, or $\left(\square\Psi\right)^2$ for the physical variable $\Psi=\lambda  -   \square\mu$. Being of a higher derivative order, such an action corresponds to two dynamical degrees of freedom, precisely like we have discussed above. They claimed \cite{Lust} that one of those can be removed by the remnant gauge freedom of the St{\"u}ckelberg trick, whatever it be in this case. It is certainly true in the sense of ambiguity of the chosen representation for the metric. However, what to make out of the second mode here? There are only very vague words, about the remnant gauge freedom again, in that paper \cite{Lust} about it. Let me stress it once more that what has happened in reality is an adverse effect of making a substitution into the action functional.

\subsection{A 3-form toy model}

Let me also use the 3-form model from the same paper \cite{Lust} for illustration of how their "manifestly covariant" approach does change the model. In terms of the dual vector, the model has got an action
\begin{equation}
\label{3form}
S=\frac12 \int d^4x \cdot (\partial_{\mu} A^{\mu})^2
\end{equation}
with the obvious equation of motion:
\begin{equation}
\label{3feq}
\partial_{\mu}  (\partial_{\alpha} A^{\alpha}) =0.
\end{equation}

In other words, the whole dynamics (\ref{3feq}) of the theory (\ref{3form}) means that
\begin{equation}
\label{3eq}
\dot A_0 - \partial_i A_i = \mathrm{const}.
\end{equation}
The spatially transverse $A^{\mathcal T}_i$ part, $\partial_i A^{\mathcal T}_i \equiv 0$, represents the obvious gauge freedom. Otherwise, we can say that the temporal component is a pure gauge one, too, while the longitudinal mode is then constrained by the equation (\ref{3eq}). Therefore, no dynamical degrees of freedom, except one half global mode.

If we do the standard St{\"u}ckelberg trick
$$A_{\mu}\longrightarrow {\tilde A}_{\mu}= A_{\mu}+\partial_{\mu}\phi,$$ 
it gives absolutely the same result (\ref{3eq}), with an extra gauge freedom of course: the action
$$S=\frac12 \int d^4x \cdot (\partial_{\mu} A^{\mu} + \square\phi)^2$$
produces the equation
$$\dot A_0 - \partial_i A_i + \square\phi= \mathrm{const}\qquad \mathrm{or} \qquad \dot {\tilde A}_0 - \partial_i {\tilde A}_i = \mathrm{const}.$$
Note that this comes from variation of the vector field. The field $\phi$, showing more derivatives, gives a weaker equation and doesn't change anything. 

At the same time, if we go the "\`a la St{\"u}ckelberg" way of Ref. \cite{Lust}
$$A_{\mu}\longrightarrow {\tilde A}^{\mu}= A^T_{\mu}+\partial_{\mu}\phi,$$
where $\partial_{\mu} {A^T}^{\mu}\equiv 0$ and therefore $\partial_{\mu} {\tilde A}^{\mu}=\square\phi$, with the action of
$$S=\frac12 \int d^4 x \left(\square\phi\right)^2$$
in terms of the scalar field only, it is radically different, for the integrations by parts in terms of $\delta\phi$ need more boundary conditions than it was in terms of a 4D-longitudinal mode of $A_{\mu}$ itself, and therefore the extremality of the action is required for a limited class of variations only, producing equations of more freedom \cite{antiV, memim}. The variation yields
$$\square^2 \phi=0 \qquad \mathrm{which\ is\ simply} \qquad \square\left(\dot {\tilde A}_0 - \partial_i {\tilde A}_i \right)=0$$
with the spatially longitudinal mode becoming a fully dynamical one. In other words, this procedure genuinely modifies the model.

The Authors of the paper \cite{Lust} correctly guessed that the problem lies in the partial gauge choice when they put ${A^T}^{\mu}$ instead of just $A^{\mu}$ into the St{\"u}ckelberg trick. However, they describe one mode coming from the remnant freedom without any proper explanation for another mode residing inside the 4th order equation. As we have seen above, this is due to incorrectness of fixing a gauge. This is a general fact. Fixing a gauge directly inside the action generically changes the theory. It wouldn't have happened if we had done it already in the equations of motion. However, the variational prescription gets changed. In this particular case, the variation $\delta A_{\mu}$ vanishing at infinity is not equivalent to $\delta A^T_{\mu}$ and $\delta\phi$ vanishing there when $ A_{\mu} = A^T_{\mu} + \partial_{\mu}\phi$. 

On one hand, fixing a gauge directly inside the action might not be a good idea. On the other hand, it also sounds as a very strange aim, to "obtain an action that is given just in terms of the gauge-invariant variables" \cite{Lust}. How could one ever do that even for electrodynamics, let alone gravity? The gauge symmetry is a very important ingredient for writing an action. The equations can be written in terms of only the field strength $F_{\mu\nu}$, however not the action itself since the action of $F_{\mu\nu}F^{\mu\nu}$ for $F_{\mu\nu}$ as a basic variable without charges would result in $F_{\mu\nu}=0$ only, with no electromagnetic waves\footnote{This is because the gauge symmetry contains derivatives. Putting more derivatives into the action functional, like in a gauge-fixed St{\"u}ckelberg trick, makes equations weaker and allows for more modes. Removing derivatives is opposite, it trivialises equations by making them stronger. A totally different example though is the coincident gauge of covariant teleparallel models \cite{meL}.}. One could of course write an action for the spatially transverse components only since those are invariant and obey a simple wave equation, but that would be a bit of cheating since it is then a different theory which totally ignores the Coulomb field and the Gau\ss{}  law.

The cosmological-perturbations-inspired "direct" approach \cite{Lust} of
$$A_i \longrightarrow {\tilde A}_i = A_i^{\mathcal T} +\partial_i\chi$$ 
also changes the model, though in a much milder way which acts only beyond perturbation theory. Indeed, the action is then
\begin{equation}
\label{Lact}
S=\frac12 \int d^4x \cdot (\dot A_0 - \bigtriangleup\chi)^2
\end{equation}
which gives the equations
\begin{equation}
\label{Leq}
\partial_0  (\partial_{\alpha} {\tilde A}^{\alpha}) =0 \qquad \mathrm{and} \qquad \bigtriangleup  (\partial_{\alpha} {\tilde A}^{\alpha}) =0,
\end{equation}
slightly more general than the original ones (\ref{3feq}). As long as we impose the condition of fields vanishing at spatial infinity, there is no change in this case.

\section{How (not) to perform the Hamiltonian analysis}

In the paper \cite{Lust}, they also confirmed their findings by using the Hamiltonian analysis. With the correct result on the number of dynamical degrees of freedom, unfortunately this analysis also modified the contents of the models, for they went for extended Hamiltonians. I would like to use this occasion to explain, see also my paper \cite{antiDirac}, that {\it unless we accurately redefine} a model with secondary first-class constraints, the Dirac's conjecture is always wrong, even in cases when it is commonly taken as correct, and therefore the secondary first class constraints should not be added to the Hamiltonian, 

The main point is that, in most interesting cases, the first class constraints are {\it not} generators of gauge transformations. Only special combinations of the primary and the secondary ones are \cite{Castellani}. At the same time, adding the secondary constraints to the Hamiltonian increases the number of arbitrary Lagrange multipliers beyond the actual amount of gauge symmetry. And this is wrong, unless we admit that some momenta lose their direct relation with the fields and grab the physical meaning of the constrained variables which then become pure gauge. 

In most expositions, this is simply swept under the rug as if only dynamical modes were physical. In electrodynamics, it then means \cite{antiDirac} that the temporal and spatially longitudinal components of the vector-potential can be gauge transformed independently of each other, thus stripping the Coulomb potentials of physical significance. In gravity, the same injustice happens \cite{antiDirac} to the scalar (and vector) perturbations of the metric which are actually observed in temperature fluctuations of CMB.

\subsection{The Hamiltonian of electrodynamics}

Let me first start from briefly summarising the situation in electrodynamics \cite{antiDirac}. For a vector field (\ref{stvec}) of any mass, the canonical momenta are $\pi_0=0$ and $\pi_i=\dot A_i - \partial_i A_0=F_{0i}$. In the massless case, it is electrodynamics in vacuum, and the total Hamiltonian density
$${\mathcal H}_T=\frac12 \pi_i^2 + \pi_i \partial_i A_0 + \frac14 F^2_{ij} + \lambda \pi_0,$$
with $\lambda$ being a Lagrange multiplier, produces then the equations of motion
$$\left\{\begin{array}{cc}
 {\dot A}_0= \lambda  \\
   {\dot A}_i = \pi_i + \partial_i A_0
\end{array}\right.
 \qquad \mathrm{and} \qquad \left\{\begin{array}{cc}
 {\dot \pi}_0=\partial_i \pi_i  \\
    {\dot \pi}_i = \partial_j F_{ji} 
\end{array}\right. 
\qquad \mathrm{with\ the\ constraint} \qquad \pi_0=0.$$
As long as we take the primary constraint as an equation rather than an initial datum, this system totally reproduces the Lagrangian dynamics of $A_{\mu}$ as well as the definition of momenta.

If we add the obvious secondary constraint $\partial_i \pi_i=0$ to the Hamiltonian density, we get the extended one
$${\mathcal H}_E=\frac12 \pi_i^2 + \pi_i \partial_i A_0 + \frac14 F^2_{ij} + \lambda \pi_0 + \tilde\lambda \partial_i \pi_i$$
with the equations
$$\left\{\begin{array}{cc}
 {\dot A}_0= \lambda  \\
   {\dot A}_i = \pi_i + \partial_i A_0+ \partial_i \tilde\lambda
\end{array}\right.
 \qquad \mathrm{and} \qquad \left\{\begin{array}{cc}
 {\dot \pi}_0=\partial_i \pi_i  \\
    {\dot \pi}_i = \partial_j F_{ji} 
\end{array}\right. 
\qquad \mathrm{with\ constraints} \qquad \pi_0=\partial_i \pi_i=0.$$
We see that the dynamics of $A_{\mu}$ has been changed. Only the transverse spatial modes are the same as they used to be. However, both the temporal and the longitudinal fields have become unphysical. Therefore, we've got twice the correct amount of gauge freedom, and totally got rid of the Gau{\ss} law. The definition of the spatial components' momenta is also lost.

What can be done is a total redefinition of the model. We say that $F_{ij}$ are still considered physical, but not the mixed components of $F_{0i}$. Instead of the latter we take $\pi_i$ as physical, forgetting its definition in terms of the fields. If in the Lagrangian equations we substitute $F_{0i}$ by $\pi_i$, it yields $\partial_i\pi_i=0$ (Gau{\ss} law in absence of charges) and $ {\dot \pi}_i = \partial_j F_{ji}$. Therefore, the extended Hamiltonian picture is then reproduced. The price to pay is abandoning the definition of canonical momenta. The electrical field then has no relation with the vector potential, and lives in the momentum space instead.

\subsection{The new life of momenta}

In a sense, what we saw about extending the Hamiltonian of electrodynamics is similar to the St{\"u}ckelberg trick. We artificially enhance the space of variables by pronouncing the momenta independent. After that we add yet another gauge symmetry by simply calling anything unphysical as long as it does not commute with any of the first-class constraints. This is a well-defined symmetry. The constraints act independently in the sense of commuting with each other. Unlike second-class ones, they cannot fully restrict a coordinate and its own momentum simultaneously. In the example of electrodynamics, only momenta are in the constraints, therefore the corresponding transformations act only on the fields not touching the momenta which can then serve as new gauge-invariant quantities being equal to their initial non-invariant definition in a particular gauge.

The same happens in general. If the constraints are of the first class, then in each canonical pair of variables they can involve only one combination and transform another one. Therefore, for each canonical pair there must exist at least one gauge-invariant combination which can then serve as a substitute for the corresponding field. And by the very definition of the new gauge symmetry, the dynamics of invariant variables are not changed by adding the secondary constraints to the Hamiltonian. Therefore, in other cases of regular structure, it should also work\footnote{See however the Ref. \cite{newa}.}. If we believe that the Hamiltonian mechanics is no less fundamental than the Lagrangian one, there is nothing wrong in considering such systems, too. I have nothing against this formal procedure as long as we honestly admit that we drastically change a model in order to get this result.

Actually, they knew that they had changed the system in the classical papers \cite{extH} "proving" the Dirac conjecture. The Lagrangian variables' behaviour was obviously changed. However, in a strange flow of thoughts, taking a Poisson bracket with any first-class constraint was voluntaristically called gauge transformation, and then only gauge-invariant variables were considered physical. For the "physical" variables the initial dynamics then gets restored, while (some) momenta lose their previous firm relation to the fields which, in turn, give up their physical meaning to the momenta. 

Unfortunately, an extended Hamiltonian is often claimed to be the only appropriate one for gauge theories, without admitting the change in the dynamical system and without explaining that important parts of physical information got transported to momenta. It often leads to rather strange ideas. Sometimes people say that only two modes of electromagnetic field are physical, ignoring the fact there is a third physical, even though constrained, mode in the Coulomb potential. It is important to remember that, in the extended Hamiltonian picture, it got moved to the sector of momenta. 

The same is for gravity. In GR, there are two dynamical modes and four more physical ones \cite{antiDirac}. If we go for an extended Hamiltonian, it is very important to remember that these constrained physical modes do not disappear but rather get rewritten in new terms, and then not only the fields but also momenta must be taken as independent and potentially physical variables. The primary constraints (the momenta of lapse and shift) are not generators of diffeomorphisms, for the coordinate changes are not about changing these functions alone. There are eight first-class constraints and only four arbitrary coordinates.

Still, I have also heard from some colleagues that there are only two physical components of the metric. In the Lagrangian, or total Hamiltonian formalism, this is certainly not true. The four constrained physical variables have got a different representation in the extended Hamiltonian description, though. However, it is very important to remember that they are there, especially if we are going to generalise the model. As one of the most trivial examples ever, we may generalise vacuum GR by adding matter contents, or just test particles. For the latter, the usual way is to look at the geodesics. And in this language, it is very important to have the full metric, in all its six physical components, and not just the two polarisations of the gravitational waves.

To sum it up, it's not that one can prove that all first class constraints correspond to gauge symmetries. We can reformulate the model, though, by artificially extending the gauge symmetry so that they all indeed generate the new gauge transformations. An important point to bear in mind is that the physical information on the constrained modes gets then transferred from the fields to (their combinations with) momenta. When studying problematic loci of the phase space or generalising a model, one should not forget about the differences.

\subsection{The 3-form example}

Now I would like to illustrate the workings of extended Hamiltonians at the toy model of Ref. \cite{Lust}. Let's start from its initial dual-vector formulation (\ref{3form}). We immediately see the canonical momenta
$$\pi_0=\dot A_0 - \partial_i A_i, \qquad \pi_i =0$$
and the total Hamiltonian
\begin{equation}
\label{3tot}
{\mathcal H}_T = \frac12 \pi_0^2 + \pi_0 \partial_i A_i + \lambda_i \pi_i
\end{equation}
with the Lagrange multipliers $\lambda_i$. The Hamiltonian equations together with the primary constraints
$$\dot A_0 = \pi_0 + \partial_i A_i, \qquad \dot A_i =\lambda_i, \qquad \dot\pi_0 = 0, \qquad \dot\pi_i=\partial_i \pi_0; \qquad \pi_i=0$$
immediately reproduce the equation (\ref{3eq}) and the definitions of momenta above. Of course, the secondary constraints are obvious from the equation for $\dot\pi_i$. In this sense, they are already in the system.

At the same time, if we add the secondary constraints of $\partial_i \pi_0 =0$ to the total Hamiltonian (\ref{3tot}), we get the so-called extended one
\begin{equation}
\label{3ext}
{\mathcal H}_E = \frac12 \pi_0^2 + \pi_0 \partial_i A_i + \lambda_i \pi_i + \tilde\lambda_i \partial_i \pi_0.
\end{equation}
The dynamical system corresponding to the new Hamiltonian (\ref{3ext}) is then totally different:
$$\dot A_0 = \pi_0 + \partial_i A_i - \partial_i \tilde\lambda_i, \qquad \dot A_i =\lambda_i, \qquad \dot\pi_0 = 0, \qquad \dot\pi_i=\partial_i \pi_0; \qquad \pi_i=0, \qquad \partial_i \pi_0=0.$$
The definition of $\pi_0$ is lost, as well as the Lagrangian equation of motion (\ref{3eq}) has vanished without a trace,  for all the vector field components have become absolutely arbitrary. The whole vector is now pure gauge, unlike it was before the addition of the secondary constraints to the Hamiltonian.

In the philosophy of proving the Dirac conjecture \cite{extH}, the role of the physical variable is now taken by $\pi_0$, instead of $\dot A_0 - \partial_i A_i$. Formally, everything is correct. The equations tell us that $\pi_0 = \mathrm{const}$. In other words, the only physical variable is indeed constant. However, we have totally lost its relations with the defining vector field. It's rather hard to take it as a reasonable step in the analysis.

As a side remark, note that here we also have another example of why one should be careful when doing the constrained Hamiltonian analysis. The three secondary constraints are all independent of each other, for they remove dependence on different spatial coordinates. However, even when taken all together, they are weaker than a single constraint of $\pi_0=0$ which is not in the system. If for counting the dynamical degrees of freedom we naively subtracted the number of all first-class constraints, then it would give us a negative number of modes. In other words, when there are only four variables, the three gauge freedoms cannot all "hit twice". 

Actually, the spatially transverse components can be taken as just totally absent from the Lagrangian, and their primary constraints simply convey this fact without hitting twice. Since we treat the space and time separately in the Hamiltonian analysis, this is not the case for the one remaining gauge freedom which does hit twice by imposing a constraint on the only physical combination of the vector components. This is analogous to what happens in the Teleparallel Equivalent of General Relativity \cite{antiDirac}. Its diffeomorphisms do hit twice, as they must, while the Lorentzian constraints simply show that the Lorentz rotations of the tetrad are not in the action functional at all, except for the boundary term of course, and those do not hit twice.

For completeness and for following the paper \cite{Lust} more closely, let me also show this analysis for the modified version (\ref{Lact}) of the model. It has removed all the spatially transverse components, though with a slight change of the physical contents of the model due to higher (spatial) derivatives in the action. We get the canonical momenta
$$\pi_0=\dot A_0 - \bigtriangleup\chi,\qquad \pi_{\chi} =0$$
and the total Hamiltonian
\begin{equation}
\label{Ltot}
{\mathcal H}_T = \frac12 \pi_0^2 + \pi_0 \bigtriangleup\chi + \lambda \pi_{\chi}.
\end{equation}
Note that the Hamiltonian equations and the primary constraint
$$\dot A_0 = \pi_0 + \bigtriangleup\chi , \qquad \dot \chi =\lambda, \qquad \dot\pi_0 = 0, \qquad \dot\pi_{\chi}=-\bigtriangleup \pi_0; \qquad \pi_{\chi}=0$$
do reproduce again both the Lagrangian equations (\ref{Leq}) and the definitions of momenta.

Adding the secondary constraint and working with the extended Hamiltonian
\begin{equation}
{\mathcal H}_E = \frac12 \pi_0^2 + \pi_0 \bigtriangleup\chi + \lambda \pi_{\chi}+ \tilde\lambda \bigtriangleup \pi_0,
\end{equation}
we end up with the modified system of equations
$$\dot A_0 = \pi_0 + \bigtriangleup\chi + \bigtriangleup\tilde\lambda  , \qquad \dot \chi =\lambda, \qquad \dot\pi_0 = 0, \qquad \dot\pi_{\chi}=-\bigtriangleup \pi_0; \qquad \pi_{\chi}=0, \qquad \bigtriangleup \pi_0=0$$
which has two arbitrary Lagrange multipliers instead of only one. The definition of $\pi_0$ is lost, though it has again become the only physical quantity, while all the Lagrangian variables have become fully free, contrary to the Lagrangian equation  (\ref{Leq}) of the model we started with.

\subsection{On the Proca fields with variable mass}

Coming back to the models with ill-defined numbers of degrees of freedom, the singular locus of the variable-mass Proca field (\ref{3ex}) is precisely $\psi=0$. The two constraints which always exist in this model, $\pi_0=0$ and $\partial_i \pi_i + 2\psi A_0=0$ get their Poisson bracket vanish there and only there. When $\psi <0$, the model has got a ghost. The traditional way of seeing it is via a St{\"u}ckelberg trick of $A_{\mu} \longrightarrow A_{\mu} + \partial_{\mu} \phi$ which makes the longitudinal mode a full-fledged ghost, for $\phi$ is then given a negative kinetic energy, or one can solve for the pair of the second-class constraints.

I am not interested in the ghosts right now, as long as they are stably there in the spectrum. Therefore, let me only briefly remind the reader about the ghost of the tachyonic vector field. For simplicity, consider just a standard Proca field (\ref{stvec}) with $m^2<0$. One easily finds two constraints, the primary $\pi_0=0$ and the secondary $A_0=-\frac{1}{m^2}\partial_i \pi_i$ ones, which can easily be solved for, and upon substitution also produce an unbounded kinetic part $\frac12 \left(\pi^2_i + \frac{1}{m^2} (\partial_i \pi_i)^2 \right)$ of the Hamiltonian. I would say, it is still an interesting question, how dangerous this instability is, for predictability of (non-linear) classical evolution. However, this is beyond the scope of this paper.

Note also that, as long as $m^2\neq 0$, the two constraints are of second class, and they are the only constraints in the system. Unlike in the case of first-class constraints, the secondary constraint can safely be added to the Hamiltonian,
$${\mathcal H}_E=\frac12 \pi_i^2 + \pi_i \partial_i A_0 + \frac14 F^2_{ij} + \frac{m^2}{2} (A^2_i - A_0^2) + \lambda \pi_0 + \tilde\lambda (\partial_i \pi_i + m^2 A_0)$$
since the Lagrange multipliers are not then free. In particular, conservation of the secondary constraint requires that $\lambda$ gives the correct value of $\dot A_0$ for making $A_{\mu}$ divergenceless, while conservation of the primary one leads then to $\tilde\lambda=0$ which also keeps the standard relation of $\pi_i$ to $\dot A_i$. 

We see that the extended Hamiltonian dynamics of massive and massless vectors are even more different from each other than their Lagrangian descriptions are. While the total Hamiltonian always keeps the definition of canonical momenta intact, the extended one loses this definition in the massless case. In this respect, when we are looking at the linearised theories, it makes no sense to go for extending the Hamiltonian. If there was an accidental gauge symmetry, it would take us even farther away from the full model. This is not surprising, of course, because we then artificially extend the accidental symmetry which is not present in the full theory.

It is also interesting to look at the Hamiltonian picture of the non-linear Proca model (\ref{exProca}) with the potential energy of $\frac14(A_{\mu}A^{\mu})^2$. We find the canonical Hamiltonian
\begin{equation}
\label{HamProca}
{\mathcal H} = \frac12 \pi^2_i + \pi_i \partial_i A_0 +\frac14 \left(F^2_{ij} + (A_{\mu}A^{\mu})^2\right),
\end{equation}
the primary constraint
\begin{equation}
\label{prc}
\Phi=\pi_0,
\end{equation}
and the secondary one
\begin{equation}
\label{scc}
\tilde\Phi = \partial_i \pi_i + A_{\mu} A^{\mu} A_0= \partial_i \pi_i +  A^3_0 - A_0 A^2_i.
\end{equation}
Generically, those (\ref{prc},\ref{scc}) are the only pair of second-class constraints. However, we immediately see that their Poisson bracket
$$\{\Phi, \tilde\Phi\}=3A^2_0 - A^2_i$$
vanishes when $3A^2_0 = A^2_i$. Since, away from $A_{\mu}= 0$, there does not seem to be any gauge freedom in the system, it must produce more constraint(s), as we have seen in the Lagrangian equations (\ref{fun1},\ref{fun2}) indeed.

What we see here was dubbed "constraint bifurcation" in the context of metric-affine generalisations of gravity \cite{Nbif, Nbif2}. These troubles are indeed very often much easier detected in the Hamiltonian approach. However, it is not then that clear how to deal with them. In the Lagrangian formalism, we do at least have a partial differential equation whose ill-posed Cauchy problem can be analysed. In Hamiltonian mechanics, the theory of non-degenerate Lagrangian systems is well-known. It is beautiful and mathematically rigorous. However, for the constrained systems, it is mostly presented rather as a recipe, which actually assumes constant ranks of the Poisson brackets' algebras of constraints. It is not clear, how to use it for calculating a number when the number is not well-defined, to start with.

\section{Discussion}

Constructing examples with scalar fields requires some conspiracy of the authors for getting pathological cases. However, the situation drastically changes when there are non-trivial internal structures in the basic fields. As we have seen above, already the simplest vector fields can produce undesired effects when faced with very natural attempts at generalising them.

The very well-known fact is that both vectors and tensors have extremely different properties depending on whether they are massive or massless, all the way up to different numbers of degrees of freedom. Therefore, unlike the case of scalars, there is no smooth limit of the mass going to zero. And there absolutely cannot be one, due to the jump in the number of dynamical modes. Any mechanisms of Vainshtein type are nothing but screening effects in particular regimes. It doesn't render them less interesting, but it doesn't make the initial value problems well-posed either, in models with an effective mass passing through the zero value.

Formally, one can make the massless equations similar to the massive ones by imposing the Lorenz gauge of $\partial_{\mu}A^{\mu}=0$. But it's not possible to do so in a model with an effective mass parametrised by a scalar field (\ref{3ex}). It would require a condition of $\partial_{\mu} (\psi A^{\mu})=0$ which then pretty much depends on how the scalar field behaves in the regime of its small values. Even more intricate catastrophes can happen if the mass depends on a more elaborate object than a mere scalar field.

As an old and pathological but rather curious example, one can recall the vector inflation \cite{vector}. Normally, vector fields decay in an expanding universe. So, we gave them a tachyonic mass. Of course, it immediately implies the problem of longitudinal ghosts \cite{vbad}, even though one can have doubts about how problematic they really are \cite{vgood}. Still, even if the inflationary behaviour can be reliably described, there must also be a moment in time when the effective mass goes through zero, for the tachyonic inflaton should probably become a normal matter component later on. This is a domain in which the model does not possess a well-defined number of degrees of freedom, and there is no surprise that it does not go smoothly \cite{vbad, vstrange}.

At the same time, there is also an outright problem in this approach which, strangely enough, was noticed only later \cite{vstrange, genvec}. Namely, the effective mass got its main contribution from a non-minimal coupling of $R A^{\mu} A_{\mu}$ type. Unless we go for Palatini formulation, the scalar curvature contains second derivatives of the metric. It means that the gravitational equations feature the second time derivatives of all the vector field components including $A_0$. In other words, all four vector field components appear to be dynamical\footnote{Actually, there are many vector fields in vector inflation. They cannot all obtain dynamical  temporal components via the gravitational dynamics only. However, one combination of those does become dynamical as can be seen via a conformal transformation to the Einstein frame \cite{vstrange}.}, and that can hardly be any stable. Moreover, the very inflationary background itself presents a strong coupling for this extra mode, for it has  $A_0=0$ and the unwanted acceleration comes in the combination of $A_0 \ddot A_0$.

To summarise the landscape of the strong coupling issues, they generically tend to appear when the constraints have a complicated non-linear structure \cite{Nbif, Nbif2}. The most obvious problematic loci are extremal points of non-linear functions of kinetic terms in the action functional, like the locus of $f^{\prime}=0$ in $f(R)$ gravity. Another common example is $f^{\prime\prime}=0$ if the argument of the function contains an essential quantity in the shape of a total (Levi-Civita) divergence. It applies both to $f(R)$ and to non-linear extensions of teleparallel gravities. And the whole field of modified teleparallel gravity is even more problematic in general, due to gauge symmetries of GR-equivalent options being broken in a very feeble manner, with rich but severely unstable sets of remnant symmetries \cite{Tissues}. Then, even without gravity, if a locus of zero mass restores a broken gauge invariance, it is a singular option. In gravity theories, it applies to massive gravity of variable mass and to symmetry-breaking non-minimal couplings of vector fields. Finally, one can intentionally produce an ill-defined Cauchy problem \cite{Vik1, Vik2} by changes of variables of a singular type, like $x \leftrightarrow x^3$, inside the action functional.

\subsection{On the $R+R^2$ and pure $R^2$ gravities again}

As we see, presence of gravity makes many pathologies more complicated, even when it concerns the usual metric approach to such theories. The recent paper \cite{Lust} considered a very simple case of pure $R^2$ gravity around its singular locus of Minkowski spacetime. Strangely enough, they present two different methods of calculating the number of degrees of freedom which yield contradictory results. They finally stick to the correct answer of no dynamical modes at all, however do not provide a proper explanation for what has happened in the alternative approach. Moreover, they then continue with an absolutely wrong statement that the absence of dynamical degrees of freedom extends beyond the linear order and to the full $R+R^2$ gravity, too.

Coming back to that, I confirm that the $f(R)=R^2$ gravity linearised around Minkowski spacetime has zero dynamical degrees of freedom \cite{Lust}, at least when taken at the face value. It happens due to an accidental gauge symmetry which obviously emerges at the point of $f^{\prime}=0$. The vanishing of tensorial polarisations is then a straightforward consequence of the fact that these modes have fully become pure gauge. What used to be a usual gravitational wave is no longer in the action at all. 

However, the scalar dynamical mode passing away has a more curious origin. It comes from disappearance of a constraint. The two Newtonian potentials in $f(R)$ gravity would normally have one combination dynamical and another combination constrained. What happens here is that the corresponding constraint has gone, to the higher orders of perturbations. Then the only remaining equation relates the d'Alembertian of a would-be dynamical mode to the Laplacian of another (\ref{redeq}). The latter would normally be constrained in itself. However, having lost the constraint, one should probably treat the initial data as just a remnant gauge freedom.

Let me discuss it at the level of the quadratic action (\ref{R2linact}) once again:
$$S=\int \sqrt{-g} d^4x \cdot R^2 = \int d^4x \cdot \left(\vphantom{\int}\left(\partial_{\mu}\partial_{\nu}-\eta_{\mu\nu}\square\right) h^{\mu\nu} \right)^2 + {\mathcal O}(h^3).$$
Fully neglecting the ${\mathcal O}(h^2)$ correction, the equations of motion are
$$\left(\partial_{\mu}\partial_{\nu}-\eta_{\mu\nu}\square\right)\left(\vphantom{\int}\left(\partial_{\alpha}\partial_{\beta}-\eta_{\alpha\beta}\square\right) h^{\alpha\beta} \right)=0.$$
We see that only one combination of metric components, $\left(\partial_{\alpha}\partial_{\beta}-\eta_{\alpha\beta}\square\right) h^{\alpha\beta} \approx R$, is predictable. All the rest is just a gauge freedom, partly from diffeomorphisms, partly an accidental one of the limit. The equations say that $R$ is a linear function of coordinates. Neglecting global degree(s) of freedom, in perturbation theory we can take it as 
$$\left(\partial_{\alpha}\partial_{\beta}-\eta_{\alpha\beta}\square\right) h^{\alpha\beta}=0.$$
Note that it does have accelerations in it. If we had a proper constraint on top of it, it would be a dynamical equation. However, missing any other information, we can say that everything is pure gauge, except the temporal component which is constrained because it can be found in terms of other components via $\bigtriangleup h_{00}$ with no need of initial data. In other words, when we face an accidental gauge symmetry, counting the modes receives a big interpretational aspect in it.

The dynamical mode(s) seen in the "covariant approach" \cite{Lust} are indeed due to the ambiguity of the spin projector formalism (\ref{strpar}), or what they called an incomplete gauge fixing. However, it can only explain one of the two dynamical modes of the 4th order equation they had got. Another mode comes from the mistake of fixing a gauge inside the action functional. Unfortunately, these problems of gauge-fixed Lagrangians are rarely discussed, though there are nice exceptions, for example in the realm of cosmological perturbations \cite{Lagos}.

To sum it up, the "pure $R^2$ gravity", i.e. $f(R)=R^2$, and the "full $R^2$ gravity", i.e. $f(R)=R+R^2$, are genuinely different from each other if to think of their behaviour around Minkowki space. This space is a singular locus for the former and is completely regular for the latter. Otherwise, they are very similar. Generically, they both have three dynamical degrees of freedom and four more physical modes which are constrained, and both have singular loci in their phase spaces, with the only difference in that it is about the points with $R=0$ for the former, and about some non-zero value of $R$ for the latter.

Moreover, even the singular behaviour of the full $R^2$ model is very similar to the pure $R^2$ case, though around another background. Indeed, with a proper cosmological constant to compensate for $f(r)$, one can take a de Sitter space of  $R=r$ with the constant $r$ such that $f^{\prime}(r)=0$. If $h_{\mu\nu}$ is the metric perturbation around the de Sitter and $\delta R = R - r$, we get
$$\delta S=\frac{f^{\prime\prime}(r)}{2} \int \sqrt{-g} d^4x \cdot (\delta R)^2 + {\mathcal O}((\delta R)^3)= \frac{f^{\prime\prime}(r)}{2} \int d^4x \cdot \left(\vphantom{\int}\left(\bigtriangledown_{\mu}\bigtriangledown_{\nu}-g_{\mu\nu}\square\right) h^{\mu\nu} \right)^2 + {\mathcal O}(h^3)$$
for the quadratic action, with all the consequences for the linearised dynamics, though with a first order time derivative of $h_{00}$ in $\delta R$ thus featuring one half a dynamical degree of freedom.

Talks of strong coupling scales \cite{Lust} are irrelevant. These effects of strong coupling which we face at $f^{\prime}=0$ are not just about an interaction being too strong for being taken into account as a small perturbation. It is about mathematical ill-posedness of an initial value problem for classical equations. Roughly speaking, it is an infinitely strong coupling. In particular, one can introduce a small non-zero cosmological constant to the pure $R^2$ gravity and study perturbations around the corresponding de Sitter solution. Those will have three dynamical modes and require the corresponding Cauchy data.  Moreover, locally the full set of exact solutions, as long as they never touch the value of $R=0$ in a domain under consideration, features the same freedom of varying the initial data. The disappearance of each and every dynamical mode is an artifact of using the sick background. Of course, it can also be seen in a purely quantum-field-theory language \cite{Karan}.

\subsection{On modified teleparallel degrees of freedom}

Alternative geometric foundations of gravity can very often be much harder to make sense of. For example, the torsion-based teleparallel models of gravity do have an extra set of Lorentz variables on top of the metric. In the current literature on modified teleparallel theories, there are many cases of ubiquitous strong coupling issues related to them. The dynamical analyses of those theories usually either ignore possible jumps in the dynamical properties or straightforwardly go to work around a singular background. As we saw above, in the example of a very simple modified gravity model, there might be very intricate reasons for the numbers changing. 

The situation in $f(\mathbb T)$ is very unclear. The available Hamiltonian analyses \cite{Ham1, Ham2, Ham3} do not agree with each other, not even in the number of degrees of freedom. In 4D, and on top of the usual two polarisations, the old claim \cite{Ham1} is that there are three new degrees of freedom (that is five in total). The analysis is not very clean, and even a much more accurate recent account \cite{Ham3} confirming this result does not offer any reliable analysis of when the jumps in degrees of freedom occur. That would require a detailed exposition of what happens with the numbers of constraints and the ranks of their Poisson brackets algebra throughout the phase space which is by far not an easy task. Another Hamiltonian analysis \cite{Ham2}, which claimed only one extra dynamical mode, had for sure committed at least one essential mistake \cite{Ham3}. 

Given this fact, it is really amazing to look at what we have seen in perturbations. Around the trivial Minkowski background of a unit matrix tetrad, there are no new modes at all, due to an obvious reason: an accidental restoration of the full local Lorentz symmetry takes place. Much more surprising is that no new dynamical modes were found around simple cosmological backgrounds either \cite{cosmo}, not even when spatial curvature was switched on \cite{cosmo2}. The local Lorentz symmetry is no longer there, but the new physical modes are constrained. At the same time, an investigation of a non-trivial tetrad solution corresponding to Minkowski metric had found one extra dynamical mode \cite{meM}, though with a somewhat restricted Cauchy data freedom similar to the example of the variable mass Proca field (\ref{3ex}). In other words, it rather confirms the results of Ref. \cite{Ham2}

One possible explanation for this strange coincidence is almost obvious \cite{menew}. The thing is that the non-trivial tetrad \cite{meM}, even though being quite non-symmetric, did preserve the condition of ${\mathbb T}=0$, while the main mistake of the Hamiltonian analysis \cite{Ham2} was in missing the contributions to Poisson brackets coming from spatial derivatives of the auxiliary scalar field corresponding to $\mathbb T$. However, this explanation is then very intriguing in itself. In almost every cosmological setting, away from bounces and the like, the torsion scalar has got a time-like gradient. Then $\mathbb T$ can be used precisely as a time variable, and the analysis of Ref. \cite{Ham2} seems to apply perfectly well. If true, it might mean that the model enjoys a preferred foliation, similar to the story of cuscuton fields \cite{cuscuton}.

\subsection{Conclusions}

Nowadays, modified gravity is a very important field of research in theoretical physics. It shows many examples of minor amendments fully destroying everything. It can be taken as either a curse or a blessing. On one hand, it is then difficult to find a reliable modified gravity model of the real Universe, and very easy to spend lots of time on phenomenological applications of an unstable approximation to a bad theory. On the other hand, it is then a very interesting quest to follow, and it allows us to gain deeper knowledge of the whole landscape of possible gravity theories including the general relativity itself.

 When exploring a new theoretical idea, we usually start from a Lagrangian density and the variational principle. It gives us very convenient tools for investigation. However, the Hamiltonian formalism is also very important for a variety of reasons. It features very nice symplectic geometry which is good for quantising if we understand anything about quantum physics. And even safely staying inside the classical framework, we happily get a first-order system of time evolution equations. It's crucial for the whole range of theoretical research, from numerical implementations if needed to rather abstract analyses of mathematical properties.

In the regular situations, the Hamiltonian formalism is relatively well understood, but unfortunately it is very often used as just a recipe from a textbook. However, especially when going for teleparallel geometries, we often encounter ill-posed systems of equations. They might very well be a dead end, but we anyway need to understand what happens with them and how. In my opinion, it is a very interesting and important task, to learn more about it and to see what the Hamiltonian analysis might tell us about such Lagrangian systems.

\end{document}